\documentclass[final,3p,times,twocolumn]{elsarticle}

\usepackage{graphics}

\usepackage{amssymb}

\journal{Solid State Communications}

\begin{document}

\begin{frontmatter}



\title{Effect of pressure on electric transport properties of carbon-doped EuB$_6$}

\address[label1]{Institute  of  Experimental  Physics,
 Slovak   Academy  of Sciences, Watsonova 47,
 040~01~Ko\v {s}ice, Slovakia}
\address[label2]{Institut f\"ur Festk\"{o}rperphysik, Technische Universit\"at Wien, Wiedner Hauptstrasse 8-10, 1040 Wien, Austria}
\address[label3]{Institute for Problems of Material Science, NASU, 252680~Kiev, 
			Ukraine}

\author[label1]{M. Batkova \corref{cor1}}
\cortext[cor1]{E-mail: batkova@saske.sk; Fax: +421 55 633 6292; Telephone: +421 55 792 2239}
\author[label1]{I. Batko} 

\author[label2]{E. Bauer} 

\author[label2]{R. T. Khan} 

\author[label3]{V. B. Filipov} 

\author[label3]{E. S. Konovalova}

\address{}

\begin{abstract}
We report about influence of external pressure on 
			electrical resistivity of EuB$_{5.99}$C$_{0.01}$, the compound believed to be intrinsically 
			inhomogeneous due to fluctuation of carbon content. 
Our results show that the low-temperature resistivity maximum shifts to lower temperature with applied 
			pressure, opposite to the behavior reported for stoichiometric EuB$_6$.
The origin of such qualitative difference we associate with the increasing			
		  volume fraction of the phase  
		  that is not compatible with ferromagnetic ordering 
		  (originating in regions with relatively higher carbon concentration) 
			with enhancing pressure. 
Our results support 
		 a recent proposition \cite{Batkova08} that carbon-rich regions   strongly influence
			magnetotransport properties of carbon-doped EuB$_6$, such as they play a role of ``spacers'', which prevent percolation of ferromagnetic phase. 
\end{abstract}


\end{frontmatter}

\section{Introduction}
\label{Intr}

EuB$_6$ orders ferromagnetically at low temperatures and 
		undergoes a metal-insulator phase transition 	\cite{Degiorgi97,Sullow98,Sullow00a}. 
Physical properties of this stoichiometric  hexaboride are thought to be governed by magnetic polarons 
		 \cite{Sullow00a,Snow01,Calderon04,Yu06} indicated by Raman scattering measurements \cite{Snow01} 
		to appear during the cooling at about 30~K.
As suggested by S\"ullow et~al.\cite{Sullow00a}, at 15.5~K %
		a spontaneous magnetization emerges accompanied by metalization,
		  and magnetic polarons begin to overlap and form a conducting, 
	ferromagnetically ordered phase that acts as a percolating, low-resistance path across the otherwise
	poorly conducting  sample.
With decreasing temperature the volume fraction of the conducting 
	FM phase expands until the sample becomes a homogeneous 
	conducting bulk ferromagnet at 12.6~K \cite{Sullow00a}.
High-pressure measurements  \cite{Cooley97} indicate that ferromagnetic (FM) order in EuB$_6$ is driven by an RKKY interaction.	
		
According to theoretical studies \cite{Shim06}, both electron doping, as well as pressure enhance
		the FM interaction between Eu atoms.
In agreement with this assumption experiments  \cite{Weill79,Guy80,Cooley97} show that 
		increasing pressure continuously 
	  reduces the resistivity of EuB$_6$ and enhances the temperature of FM ordering.   
		 
Because of the very low number of intrinsic charge carriers
		($\sim$10$^{20}$~cm$^{-3}$) \cite{Aronson99},
		even a slight increase in the concentration of conduction electrons 
		 drastically modifies electric and magnetic properties of this compound \cite{Batkova08,Kasaya78,Molnar81}. 
Substitution of boron by carbon increases the number of conduction electrons in the system. 
It seems that the prevailing effect of the additional electrons 
	 	(above a certain concentration) is to produce antiferromagnetic 
		exchange that competes with the initial FM interaction \cite{Tarascon81}.
As shown by neutron diffraction studies \cite{Tarascon81}, the predominant
		FM ordering in stoichiometric EuB$_6$ changes with increasing 
		carbon content through a mixture of ferromagnetic and helimagnetic domains into a purely 
		antiferromagnetic state in EuB$_{6-x}$C$_x$ with $x\geq0.125$ \cite{Tarascon81,Kasaya78}.
The helimagnetic domains are formed in carbon-rich regions, which are characterized by higher 
		carrier density than the surrounding FM matrix \cite{Tarascon81}.
Different types of magnetic order are a consequence of a distinct impact of the 
	 RKKY interaction due to distinct carrier 	densities \cite{Tarascon81}.
The presence of the helimagnetic domains is also believed to be
		responsible for an additional scattering term in the electrical resistivity 
		below the bulk FM phase transition \cite{Batkova08,Batko95}.

Recently, an anomalously large negative  magnetoresistance
		of EuB$_{5.99}$C$_{0.01}$ was reported \cite{Batkova08} and attributed to 
	 a local fluctuation in carbon content.	
The proposed new scenario \cite{Batkova08} on the role of regions with increased carbon concentration 
		 has allowed to explain huge differences between  the
		 $\rho(T)$ characteristics of pure and  carbon-doped EuB$_{6}$.
As follows from this scenario \cite{Batkova08}, 
		regions with charge-carrier concentration exceeding a certain threshold value, 
		 represented by the regions with correspondingly higher carbon concentration,  
		are not compatible with the existence of magnetic polarons (and with FM state in general). 
Thus, in the temperature region close above the temperature of the bulk FM ordering, $T_c = $ 4.3~K, 
		these regions act as spacers that	prevent magnetic polarons to link, 
		 to form FM clusters, and eventually to percolate \cite{Batkova08}.
As a result, the percolation and the bulk FM state 
		occur at lower temperature. 
According to this scenario the principal differences 
		between the pure EuB$_6$ and EuB$_{5.99}$C$_{0.01}$ can be summarized as follows \cite{Batkova08}.
(i) While the paramagnetic state in EuB$_6$ is homogeneous, 
		the paramagnetic state in  EuB$_{5.99}$C$_{0.01}$ has to be treated as inhomogeneous, 
 		containing regions with increased carbon content and with  correspondingly
		higher electrical conductance in comparison to the remaining matrix. 
(ii) The magnetic polaron phase in EuB$_6$ can be treated as two-component system,
		consisting of highly conducting FM phase represented by MPs and poorly conducting paramagnetic matrix.
	  This phase in  EuB$_{5.99}$C$_{0.01}$ does have three components at least: 	
	  highly conducting FM phase represented by MPs, regions with lower carbon content, and carbon-rich domains. 
The latter, incompatible with the existence of MPs due to too high charge carriers concentration, 
		 work as ``spacers", preventing MPs to link and to form a conductive path across the sample.
(iii) Finally, in the magnetically ordered state EuB$_6$ is a homogeneous ferromagnet, while 
	EuB$_{5.99}$C$_{0.01}$ can be treated as two component system consisting of 
	FM matrix and HM domains formed in the carbon-rich regions. 
An important consequence of this 
		scenario from an application point of view is that it might show a route 
	 	for optimization of the magnetoresistive  properties also in other 
		spatially inhomogeneous systems with magnetic polarons, or with FM phase in general \cite{Batkova08}.
The aim of this paper is to give another experimental support that regions 
		with increased charge-carrier concentration, not compatible with FM ordering,
		are responsible for huge differences between  pure and  carbon-doped EuB$_{6}$.

\section{Material and Methods}
	
We studied single crystal EuB$_{5.99}$C$_{0.01}$ with $T_c =4.3$~K, grown by the zone-floating method. (For more details see Ref. 1.) 
The electrical resistivity was measured 
		by a standard DC four-probe method.
Hydrostatic pressure conditions up to 10.3 kbar were generated by a piston-cylinder cell (liquid pressure cell).
The absolute value of the pressure was
		determined from the superconducting transition of Pb.

\section{Results and Discussion}	

 \begin{figure}[!b]
\resizebox{1.1\columnwidth}{!}{%
  				\includegraphics{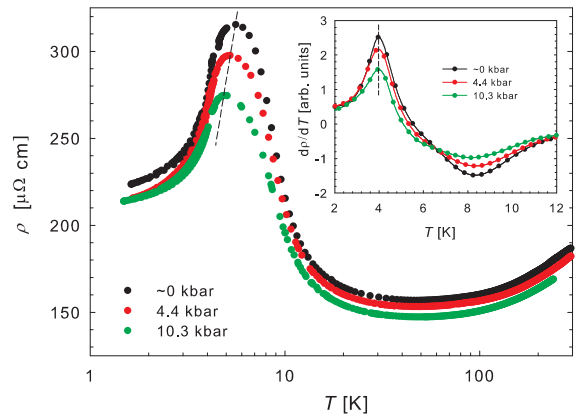}
  				}
\caption{External-pressure influence on the $\rho(T)$ and d$\rho/$d$T(T)$ dependences of EuB$_{5.99}$C$_{0.01}$.}
\label{RTplusdRdT}
\end{figure}	 	

We measured temperature dependence of the electrical resistivity between 1.6 and 300~K 
			for three distinct values of the applied pressure: $\sim$0, 4.4 and 10.3~kbar.
As it is evident from Fig. \ref{RTplusdRdT}, increasing pressure has two kinds of effect on the temperature dependence 
			of the resistivity: (i) decrease of the overall resistivity 
			in the whole temperature interval investigated,
			and (ii) a shift of the resistivity maximum towards lower temperature. 
The former one, qualitatively similar to the observation in pure EuB$_6$, 
		 can be simply ascribed to an increased 
		concentration of charge carriers due to applied pressure.
The latter one, however, shows an opposite tendency compared to pure EuB$_6$ \cite{Weill79, Cooley97}, where  
			 the (low temperature) resistivity maximum
			increases upon pressure due to 
			enhanced FM interactions caused by the increased electron concentration.  
A qualitative difference reflects itself also in a position of the maximum in d$\rho/$d$T(T)$ 
			(see inset in Fig. 			\ref{RTplusdRdT}). 
While the maximum in d$\rho/$d$T(T)$ for pure EuB$_6$ 
			shifts to higher temperatures with pressure increase \cite{Cooley97}, 
			in the case of EuB$_{5.99}$C$_{0.01}$ it does not change the position on temperature scale with respect
			to the experimental error. 
			
Especially the \emph{opposite} effect of pressure on the position of the  peak in $\rho(T)$ 
			indicates that another phenomena than those governing electrical transport properties 
			of pure EuB$_6$ should be dominating in  EuB$_{5.99}$C$_{0.01}$.
This seems to be a consequence of
			 local fluctuations in carbon content and corresponding local fluctuations
			 in charge-carrier concentration, whereas the latter leads to the magnetic phase separation  \cite{Batkova08}.
Taking into account that concentration of charge carriers is a primary parameter governing a position of the boundary between FM and 				non-FM regions, an increase in charge-carrier concentration due to applied external pressure 
		   shifts this boundary to the zone with lower carbon concentration. 
Such as spacers are defined as regions not compatible with FM ordering, 
			external pressure in fact increases their volume and space distribution.
As a consequence,  conditions for a mutual interconnection of higly conductive  FM regions 
		(resulting from an overlap of magnetic polarons)  close above $T_c$
		become unfavourable with pressure increase.
Thus, the  transition from a metal-like transport 
			(d$\rho/$d$T > 0$) to a transport regime
			with a predominant semiconducting, 
			nonferromagnetic phase 	(d$\rho/$d$T < 0$),	shifts towards lower temperature.		
In such way the scenario \cite{Batkova08} on the spacing role of the charge-carrier-rich regions in carbon-doped EuB$_{6}$
 	  yields a consistent explanation of the effect of external pressure 
		on the electrical resistivity of EuB$_{5.99}$C$_{0.01}$.

	 		 \begin{figure}[!b]
\resizebox{1.0\columnwidth}{!}{%
  				\includegraphics{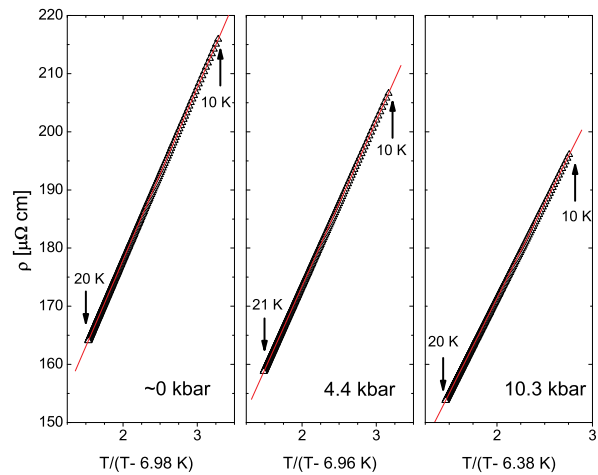}
  				}
\caption{Influence of external pressure on the $\rho(T)$ dependence.}
\label{SpinDis}
\end{figure}	 

Fig.~\ref{SpinDis} illustrates how external pressure influences the temperature dependence 
		 of the resistivity of EuB$_{5.99}$C$_{0.01}$ in the paramagnetic state, focusing on
		  the temperature between $\sim$10~K and $\sim$20~K.
As demonstrated previously \cite{Batkova08,Batkova06},  a dominant contribution to 
		the temperature dependence of the resistivity 
			 in this  temperature range 
		  arises from the spin disorder scattering;  
		the $\rho(T)$ dependence 
		can be satisfactorily analyzed using the formula 
		$\rho(T)=  A . T/(T-\theta_C)+ \rho_d$, where A is a constant, parameter $\theta_C$ is the paramagnetic 
		Curie temperature, and $\rho_d$ represents a contribution due to temperature 
		independent scattering on defects.
As can be seen  
		from Fig.~\ref{SpinDis}, 
		 external pressure lower than 4.4 kbar practically does not 
		influence the $\theta_C$ ($\theta_C = 6.98\pm0.05$~K and $6.96\pm0.05$~K 
		for $\sim 0$ and $4.4$~kbar, respectively).
At higher pressure  
		  the paramagnetic Curie temperature slightly decreases 
		  ($\theta_C = 6.38\pm0.05$~K), 
		  which could be also associated with the expansion of non-FM regions (spacers)
		  and strengthened antiferromagnetic interactions
		  inside them, consistently with the  above mentioned scenario. 

Of course, application of the scenario on the role of spacers
		requires to take into account the following aspects/limitations. 
(i) The volume fraction of the  phase  not compatible with FM ordering should be comparable to
	  the volume fraction of the vast phase, where MPs can exist.
Too high volume fraction of the phase not compatible with FM ordering would prevent the percolation process,
		thus  only islands of FM phase would exist in the system and the percolation would not occur at all; on the other hand, too low 
		 fraction of these regions would not have any important impact on the percolation process. 	 	  
(ii) Characteristic dimensions of regions  not compatible with FM ordering 
	  should be comparable with dimensions of MPs at temperatures close above $T_c$. 
If these regions were much larger than MPs, 
		theys would only represent  some "restricted regions", but could not effectively play the role of spacers,
		such as MPs could easily percolate in the remaining volume of the material.
On the other hand, if these regions are too small, they would not effectively prevent the percolation/overlap of significantly larger MPs. 

Applying the above mentioned criteria on the case of the investigated EuB$_{5.99}$C$_{0.01}$ 
		leads to the conclusion that  
		 the volume fraction of the charge-carrier-rich regions should be comparable to the vast material 
		 and characteristic dimensions of these regions should be of the same range as MPs 
		 close above $T_c$.
Unfortunately, we have no direct evidence on the volume of regions not compatible with FM state 
			in the investigated compound.
More light into the problem should be brought e.g.  by detailed neutron
			diffraction studies.) 
However, we strongly suppose a comparable volume fraction of the charge-carrier-rich regions
		 and the vast material.
(Just for {\em illustration} purposes, a consideration of the 
		system as a mixture 
		of EuB$_{5.995}$C$_{0.005}$ and  EuB$_{5.985}$C$_{0.015}$ phase would yield a volume ratio 1:1;
		 a mixture of EuB$_{5.97}$C$_{0.03}$ or EuB$_{5.95}$C$_{0.05}$, as the carbon-rich phase 
		 and EuB$_6$ as the carbon-poor one
		 would provide the ratio 2:1 or 4:1, respectively.)
A rough estimation of an expected size of the spacers and MPs in the studied system 
		can be performed as follows.   
Estimation of the radius of MPs in EuB$_6$ close above $T_c$ from Raman scattering studies \cite{Snow01}
		 yields   a value $R \approx 2a$, where $a=4.19$~\AA~  is the lattice parameter of EuB$_6$ 
		close above the $T_c=12$~K \cite{Tarascon81}.		
Thus, an estimated size/diameter of MPs in EuB$_6$ close above the $T_c$ is $2R\approx 16.76$~\AA. 
According to the neutron scattering studies reported by Tarascon et al., 
			the size of incoherent regions present in EuB$_{5.95}$C$_{0.05}$ 
	  due to the fluctuation in carbon content is about 50~\AA~  \cite{Tarascon81}.
Supposing that carbon-rich regions in EuB$_{5.99}$C$_{0.01}$ have the  same carbon concentration 
				but adequately reduced size when comparing with EuB$_{5.95}$C$_{0.05}$, the estimated size/diameter of the spacers
		in the tudied compound should be $\approx 29.2$~\AA, what is comparable with the above mentioned size of MPs in EuB$_6$. 
According to this {\em rough} estimation EuB$_{5.99}$C$_{0.01}$ can be  treated within
		the limits of the scenario of spacers \cite{Batkova08}.

The above presented results and related discussion shed new light 
		on an understanding of the very intriguing behavior of EuB$_{5.99}$C$_{0.01}$ system and allow to explain 
		qualitatively different effect of pressure 
		on the electrical characteristics		of  pure and carbon-doped EuB$_6$. 
As stated previously \cite{Batkova08}	and 
		discussed in this paper, a major difference between the two systems is represented by
		the presence of the spacers, i. e. regions not compatible with FM order due to high carrier concentration, which
		prevent FM phase to percolate	and to form an electrically conductive path along 
		the	carbon doped sample. 
It can be therefore concluded that both, the observed shift of the resistivity maximum to lower temperature, as well as 	
		a decrease of the paramagnetic Curie temperature upon increasing pressure, 
	  can be attributed to the expansion of  the spacers upon increasing pressure. 
Thus the results confirm 
			that the dominant effect of substitution of boron by carbon in EuB$_{5.99}$C$_{0.01}$.
		is an inhomogeneous volume distribution of carbon, 	causing magnetic phase separation at low temperatures,
		instead of (an homogeneous) increase of the charge-carrier concentration.
		
\section*{Acknowledgements}
 This work was supported by 
		the Slovak Scientific Agency (Project VEGA 2/0133/09), by the Slovak Research 
		and Development Agency (APVV-0346-07) and by the Austrian FWF P18054. 
	 M.B. acknowledges the support of the European Science Foundation (COST-STSM-P16-639).

\end{document}